# Systematic Refinement of Canongia Lopes-Pádua Force Field for Pyrrolidinium-Based Ionic Liquids

Vitaly V. Chaban[1] and Iuliia V. Voroshylova[2]

[1] Instituto de Ciência e Tecnologia, Universidade Federal de São Paulo, 12247-014, São José dos Campos, SP, Brazil

[2] CIQ/REQUIMTE –Department of Chemistry and Biochemistry, Faculty of Sciences, University of Porto, Rua do Campo Alegre 687, 4169-007, Porto, Portugal

**Abstract**. Reliable force field (FF) is a central issue in successful prediction of physical chemical properties via computer simulations. While Canongia Lopes—Pádua (CL&P) FF provides good to excellent thermodynamics and structure of pure room-temperature ionic liquids (RTILs), it suffers from drastically and systematically underestimated ionic motion. This occurs due to neglected partial electron transfer from the anion to the cation, resulting in unphysically small simulated self-diffusion and conductivity and high shear viscosities. We report a systematic refinement of the CL&P FF for six pyrrolidinium-based RTILs (1-N-butyl-1-methylpyrrolidinium dicyanamide, triflate, bis(fluorosulfonyl)imide, bis(trifluoromethanesulfonyl)imide, tetrafluoroborate, chloride). The elaborated procedure accounts for specific cation-anion interactions in the liquid phase. Once these interactions are described effectively, experimentally determined transport properties can be reproduced with an acceptable accuracy. Together with the original CL&P parameters, our force field fosters computational investigation of ionic liquids. In addition, the reported results shed more light on the chemical nature of cation-anion binding in various families of RTILs.

**Introduction**

Computer simulation has become a mighty and widely recognized tool to predict physical chemical properties of room-temperature ionic liquids (RTILs). A basic search performed in the ISI Web of Knowledge database on November 7, 2014 reveals that 373 papers were devoted to computer simulations of RTILs only during 2014. Compare this number with a total number of papers devoted to ionic liquids, 7072, during 2014. Hereby, 5.3 percent constitute a very significant portion, provided that an average computer simulation contribution is more intellectually expensive than a direct physical chemical experiment employing an established methodology. Most numerical simulations report atomistic-resolution molecular dynamics, based on pairwise pre-parameterized interaction potential functions. A smaller number of works employ more sophisticated interaction potentials, Monte Carlo techniques, zero-temperature electronic structure studies, ab initio molecular dynamics simulations, coarse-grained representations of interacting sites, etc.

Reliable sampling of phase space and correct representation of interatomic binding functions (known as force field, FF) are a central issue in successful prediction of physical chemical properties via various sorts of computer simulation techniques. Several developments for the state-of-the-art atomistic description of ionic liquids have been introduced over the recent years. Canongia Lopes and Pádua have essentially pioneered force field development for ionic liquids. A large variety of RTILs have been parameterized within the framework of their well-established methodology (CL&P). The developed models are internally consistent, transferrable, and compatible providing a good to excellent accuracy of the most simulated thermodynamic and structure properties at room conditions and somewhat elevated temperatures. The major advantage of the CL&P FF over other FFs is its availability for a large number of ionic species.

Another comprehensive contribution to the RTIL simulations has been delivered by Sambasivarao and Acevedo. Sixty eight unique combinations of room temperature ILs featuring 1-alkyl-3-methylimidazolium, N-alkylpyridinium, choline cations, supplemented by an extensive set of anions, have been parameterized using harmonic bonded forces and additive pairwise potential functions. The developed models have been tested using the Metropolis Monte Carlo simulations. In turn, Borodin has applied a Thole oscillator model to produce polarizable ions for a significant number of ILs. A set of simulated thermodynamic (density, heat of vaporization) and transport properties (self-diffusion coefficients, shear viscosity, ionic conductivity) appear in a reasonable agreement with the available experimental results. We have previously made certain contribution to the FF development and refinement by coupling electronic structure and molecular mechanical descriptions of inter-ionic interactions.

The present work further elaborates/justifies an approach to develop efficient atomistic force field models for RTILs under a simple approximation of additive non-bonded interactions. Starting from the FF for isolated ions in vacuum (CL&P), we consider two major corrections: (1) for electron transfer effects between the cation and the anion and (2) a correction for cation-anion atom-atom closest-approach distances. Both corrections are driven by the results of hybrid density functional theory (DFT) for ion clusters. We apply our methodology to six RTILs representing the pyrrolidinium family: (1) 1-N-butyl-1-methylpyrrolidinium tetrafluoroborate [PY14][BF$_4$]; (2) 1-N-butyl-1-methylpyrrolidinium bis(trifluoromethanesulfonyl)imide [PY14][TFSI]; (3) 1-N-butyl-1-methylpyrrolidinium dicyanamide [PY14][DCA]; (4) 1-N-butyl-1-methylpyrrolidinium bis(fluorosulfonyl)imide [PY14][FSI]; (5) 1-N-butyl-1-methylpyrrolidinium triflate [PY14][TF]; (6) 1-N-butyl-1-methylpyrrolidinium chloride [PY14][Cl]. Figure 1 introduces ion aggregates (two ion pairs), which constitute primary building blocks of the investigated RTILs. Certain experimental data for four of these RTILs are available (see Table S1). These data can be used to evaluate the accuracy of the derived FF. We have not found any experimental properties for [PY14][BF$_4$] and [PY14][Cl]. Experimental efforts to address transport properties of the two mentioned RTILs are very welcome.

**Methods and Tools**

Cation-anion specific interactions were investigated using the hybrid electronic density theory (Figure 1). We describe an electron density using the recently proposed high-quality hybrid exchange-correlation functional, omega B97X-D. The exchange energy is combined with the exact energy from the Hartree-Fock theory. Empirical atom-atom dispersion correction is included. According to Chai and Head-Gordon, the omega B97X-D functional simultaneously yields an improved accuracy for thermochemistry, electron kinetics, and non-covalent interactions. We construct the wave function using a comprehensive correlation-consistent basis set proposed by Dunning and coworkers. Based on our preliminary tests, triple-zeta basis set from this family, augmented with the diffuse functions (aug-cc-pVTZ), provides an accurate description of the electronic density and energy levels. Certain known defects, such as excessive valence electron delocalization, widely observed in case of pure DFT methods, were not reported for the presently selected method and basis set. Furthermore, we investigate how electron transfer between the anion and the cation is impacted by the number of ion pairs. We test ion clusters containing 1, 2, 3, and 4 ions pairs (i.e. up to 8 ions). Since the resulting clusters are significantly large in terms of electrons, we employed a smaller basis set, 6-31G(d) for this particular benchmark (Figure 2). Electrostatic potential (ESP), derived from the electronic wave function, was fitted using a set of point charges localized on each atom of the cation and anion, including hydrogen atoms. In case of ion pairs (i.e. neutral systems), an additional constraint was imposed during the fitting procedure to rigorously reproduce dipole moment of the ion aggregate. Atomic radii were assigned on the basis of the CHELPG scheme. All electronic structure calculations and geometry optimizations of ionic clusters were performed using the GAUSSIAN 09 (revision D) program.

Table 1. Electrostatic point charges condensed on the non-hydrogen atoms of *N*-1-ethyl-1-methylpyrrolidinium cation. The charged were assigned according to the CHELPG scheme. An isolated cation state is compared to ion pairs containing various anions. The electronic structure calculations for all systems were conducted at the omega B97XD/aug-cc-pVTZ level of theory

| Atom | 1-ethyl-1-methylpyrrolidinium$^+$ + anion |
|---|---|

|         | isolated | [BF$_4$]$^-$ | [Cl]$^-$ | [DCA]$^-$ | [FSI]$^-$ | [TF]$^-$ | [TFSI]$^-$ |
|---------|----------|--------------|----------|-----------|-----------|----------|------------|
| C-1 (ring) | +0.19 | +0.20 | +0.18 | +0.12 | +0.20 | +0.20 | +0.24 |
| C-2 (ring) | +0.04 | 0.00 | -0.01 | +0.05 | -0.01 | -0.01 | -0.06 |
| C-3 (ring) | +0.15 | +0.09 | +0.08 | +0.11 | +0.10 | +0.09 | +0.14 |
| C-4 (ring) | +0.10 | +0.12 | +0.10 | +0.06 | +0.09 | +0.11 | +0.07 |
| N-5 | +0.12 | +0.13 | +0.19 | +0.08 | +0.18 | +0.13 | +0.09 |
| C-6 (CH$_3$) | +0.18 | +0.07 | +0.05 | +0.09 | +0.08 | +0.08 | +0.11 |
| C-7 (CH$_2$) | +0.14 | +0.24 | +0.19 | +0.28 | +0.23 | +0.25 | +0.28 |
| C-8 (CH$_3$) | +0.08 | -0.01 | -0.01 | -0.02 | +0.01 | 0.00 | -0.01 |
| Total | +1.00 | +0.84 | +0.77 | +0.77 | +0.88 | +0.88 | +0.86 |

The thermodynamics (heat of vaporization, $\Delta H_{vap}$, mass density, $d$) and transport properties (self-diffusion coefficients, $D_{\pm}$, shear viscosity, $\eta$) have been obtained using equilibrium and non-equilibrium molecular dynamics (MD) simulations, as detailed below. The basic list of the simulated systems is provided in Table 2. The RTIL ions were placed in cubic periodic MD boxes, whose densities were calculated to correspond to ambient pressure at the requested temperature (298, 313, 338, 353 K). The simulated temperatures were selected in view of the currently available experimental data and primary interest for existing and prospective applications.

The first 10 ns of recorded trajectory for every system were regarded as equilibration, whereas subsequent trajectory part was used to sample the phase space and derive physical chemical properties of interest. See Table 2 for simulation durations. Note that each system at each temperature underwent non-equilibrium simulation (for shear viscosity) and equilibrium simulation (for all other properties). Cartesian coordinates were saved every 1 ps and data on energy dissipation (for viscosity calculation) were saved every 0.01 ps. More frequent saving of trajectory components was preliminarily tested, but no systematic accuracy improvement was found. Self-diffusion coefficients were computed from mean-square displacements of atomic positions versus time. Prior to these calculations, the trajectories must be pre-processed to remove information about periodic boundary conditions. Shear viscosity was calculated using cosine-shape acceleration of all ions simultaneously. The mathematical foundation of this method and its applications to molecular dynamics simulations were, in great details, discussed by Hess. The first five nanoseconds of the simulation were used for an accelerated ionic flow to be established. The subsequent 45 ns were used for the viscosity calculation. The standard error of each viscosity value was evaluated from statistical processing of 0-15, 15-30, and 30-45 ns trajectory fractions. Hereby, these trajectory fractions were treated as independent measurements in statistical analysis.

Table 2. The list of systems simulated in the present work. Each ionic composition was simulated at 298, 313, 338, and 353 K (i.e. four MD systems per ionic composition). The quantity of ion pairs per RTIL was selected with respect to the cation and the anion sizes. The durations of trajectories for various RTILs were selected in a way so that computational cost per system was roughly equal in all cases. Note that required sampling strongly depends on the system temperature and, ultimately, on the average self-diffusion constant in the respective system

| # | RTIL | Number of ion pairs | Number of interaction centers | Sampling times, ns | |
|---|---|---|---|---|---|
| | | | | Equilibrium[1] | Non-equilibrium[2] |
| 1-4 | [PY14][TFSI] | 125 | 5625 | 20 | 50 |
| 5-8 | [PY14][$BF_4$] | 150 | 5250 | 30 | 70 |
| 9-12 | [PY14][DCA] | 150 | 5250 | 20 | 50 |
| 13-16 | [PY14][FSI] | 125 | 4875 | 25 | 50 |
| 17-20 | [PY14][TF] | 125 | 4750 | 20 | 50 |
| 21-24 | [PY14][Cl] | 175 | 5425 | 30 | 70 |

[1] Equilibrium MD simulation was used to sample mass density, heat of vaporization, and self-diffusion coefficients of the cation and the anion.
[2] Non-equilibrium MD simulation was used for shear viscosity calculation as described above.

All systems were simulated in the constant-pressure constant-temperature ensemble. The equations-of-motion were propagated with a time-step of 1.0 fs. The electrostatic interactions were simulated using direct Coulomb law up to 1.3 nm of separation between the interaction sites. The electrostatic interactions beyond 1.3 nm were accounted for by computationally efficient Particle-Mesh-Ewald (PME) method. It is important to use the PME method in case of ionic systems, since electrostatic energy beyond the cut-off usually contributes 40-60% of total electrostatic energy. We used a smaller real-space cut-off, as compared to the original work of Canongia Lopes—Pádua. This methodological adjustment does not change the accuracy noticeably, according to our preliminary assessment, but allows to speed up the MD calculations. The Lennard-Jones-12-6 interactions were smoothly brought down to zero from 1.1 to 1.2 nm using the classical shifted force technique. The constant temperature (298, 313, 338, 353 K) was maintained by the Bussi-Donadio-Parrinello velocity rescaling thermostat (with a time constant of 0.5 ps), which provides a correct velocity distribution for a statistical mechanical ensemble. The constant pressure was maintained by Parrinello-Rahman barostat with a time constant of 4.0 ps and a compressibility constant of $4.5 \times 10^{-5}$ bar$^{-1}$. All molecular dynamics trajectories were propagated using the GROMACS simulation suite. Analysis of properties was done using the the in-home tools and supplementary utilities distributed with the GROMACS package, where possible.

**Force Field Derivation**

1-ethyl-1-methylpyrrolidinium cation was used instead of 1-N-butyl-1-methylpyrrolidinium cation for electronic structure calculations. The reason for such a solution is that electrostatically neutral and nonpolar -CH$_2$-CH$_3$ fragment does not participate in charge transfer between the anion and the cation. Avoiding -CH$_2$-CH$_3$ significantly decreases computational cost of the system and hastens convergence of self-consistent field procedures.

Table 1 lists point charges obtained through fitting of electrostatic potential at the surface of the isolated 1-ethyl-1-methylpyrrolidinium cation and six neutral ion pairs containing 1-ethyl-1-methylpyrrolidinium cation and various anions. The point charges on the hydrogen atoms were summed up with the charge on the carbon atom for simplicity of representation. All ion pairs exhibit a significant electron delocalization, indicating – among other observations – strong cation-anion binding in these configurations. The total charges of the respective anions are the same as of 1-ethyl-1-methylpyrrolidinium, but with an opposite sign. Cation-anion binding breaks symmetries of both isolated cation and anion. The point charges on the same heavy atoms of the cation (carbon, nitrogen) differ very significantly when the cation is paired with different anions. Compare, q(N)=+0.12e in the isolated [PY12], q(N)=+0.19e in [PY12][Cl], and q(N)=+0.08e in [PY12][DCA]. Note that these calculations do not account for thermal motion effects. Thermal motion presumably brings even larger fluctuations of point charges. Since the charge of C-8 (CH$_3$ radical, which belongs to the ethyl chain, Table 1) appears nearly zero with all anions, we disregard it in the further calculations of larger ion clusters (Figure 1).

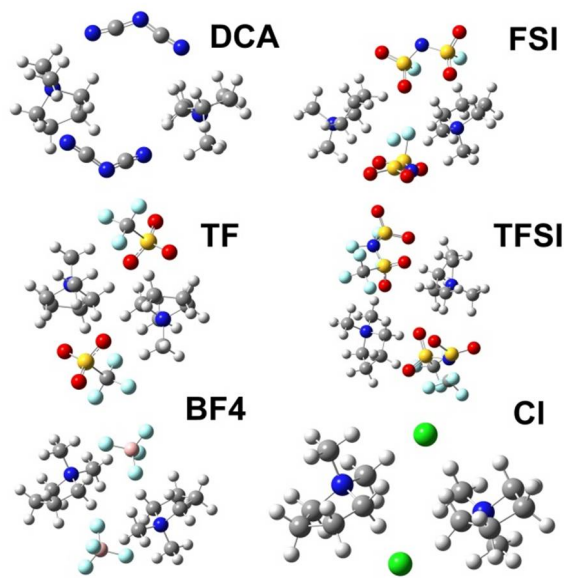

Figure 1. Optimized geometries for the isolated clusters composed of four RTIL ions each: two 1-ethyl-1-methylpyrrolidinium cations and two respective anions. 1-ethyl-1-methylpyrrolidinium cations are used in ab initio calculations instead of 1-N-butyl-1-methylpyrrolidinium cation.

The total charge on the ion can, in principle, depend on the size of ion cluster. It is a probable situation in compounds, where more than one coordination center per molecule (ion) is present. In such cases, a single counter-ion does not occupy all possible sites, which are realized in the condensed phase. Hence, charge transfer may be described partially, not completely. Figure 2 depicts optimized geometries of [PY12][DCA] clusters containing 2-8 RTIL ions (1-4 ion pairs). Figure 3 summarizes total charge on the [PY12] cation in its clusters with [DCA]$^-$ and [Cl]$^-$, as a function of the cluster size. Addition of the second ion pair brings insignificant decrease of the total charge, ca. 1%, while addition of more ions does not exhibit a systematic impact. It should be noted that electrostatic potential is well defined only at the surface of the ion cluster; therefore, further addition of ions does not necessarily correspond to a more realistic description of point charges. In turn, implicit solvent models may capture polarization effect of the surrounding solvent, but does not describe an electron transfer phenomenon (since the solvent is implicit). It is questionable whether implementation of periodic boundary conditions

(PBC) is more beneficial in the present case, since PBC still do not represent a long-range structure of RTIL. Due to small size of the unit cell, reproduction of experimental density (whether in the constant-volume ensemble or in the constant-pressure ensemble) appears problematic.

Since the total charge of an ion does not change significantly in larger clusters, both the result for one ion pair and two ion pairs can be used. In this work, we chose the result for two ion pairs (four ions), averaged to two decimal digits, to be used for the force field derivation. It should be noted that various DFT functionals provide reasonably good agreement. Compare, in the case of $[PY12]_2[DCA]_2$, PBE provides ±0.69e/ion, B97D provides ±0.70e/ion, M06L provides ±0.72e/ion, BLYP provides ±0.70e/ion. Although the difference is somewhat larger than the difference between various number of ion pairs, it can be considered sufficiently small in the context of force field derivation. We used the charges from the PBE functional for the force field derivation. This choice is arbitrary.

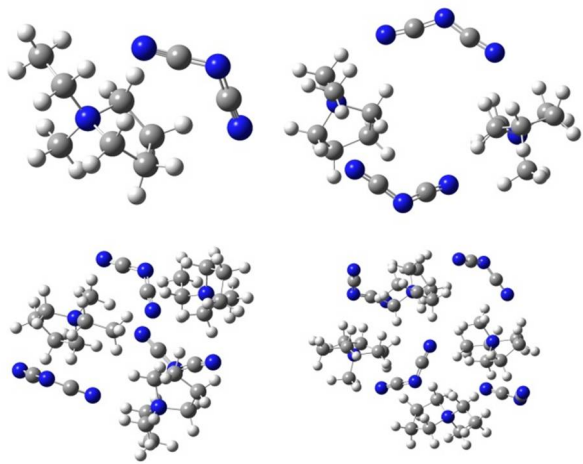

Figure 2. Optimized ionic configurations of $[PY12]_n[DCA]_n$ for n=1, 2, 3, 4.

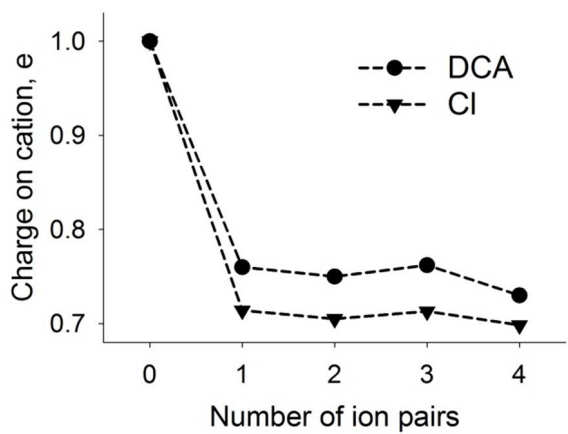

Figure 3. Electrostatic (CHELPG) charge localized on the [PY12] cation in the $[PY12]_n[DCA]_n$ and $[PY12]_n[Cl]_n$, n = 1, 2, 3, 4. The depicted charge was computed as a sum of all point electrostatic charges on the [PY12] atoms, including hydrogen atoms. The case of zero ion pairs corresponds to an isolated cation. Its charge equals to unity according to the number of electrons in the considered nuclei-electron system.

The set of immediate charges, such as those summarized in Table 1, cannot be directly transferred to the classical force field. All of these point charges are small. They fluctuate drastically upon thermal motion of RTIL. Instead, the symmetry of charge distribution (CL&P FF) can be preserved, whereas an electrostatic potential can be decreased through uniform scaling of all charges of the pyrrolidine ring of the cation and the anion. This solution was successfully applied in our previous works. The so-called scaling factor equals to the total charge on an ion suggested by DFT. Remember, that electrostatically neutral moiety of the hydrocarbon chain, $-CH_2-CH_3$, does not participate in charge transfer and, hence, charge scaling is not necessary for it. We implement uniform scaling factors for [PY14][DCA], [PY14][FSI], [PY14][TF], $[PY14][BF_4]$, and [PY14][Cl]. It could have been also applied with reasonable success to [PY14][TFSI], as illustrated before. However, in view of two sulfur atoms in [TFSI]$^-$, we attempted to account for their different polarizability by transferring charges from an ab initio calculation of $[PY12]_2[TFSI]_2$. Direct transfer of the ESP point charges from the ab initio calculation to the Coulomb potential is not doable, because formation of an ion pair breaks an original symmetry of isolated ions. That is, symmetric atoms of sulfur, oxygen, carbon, and fluorine in the [TFSI]⁻ anion obtain significantly alternating charges. Being averaged to the first decimal digit, the point charges are -0.4e (N), +0.6e (S), -0.4e (O), +0.3e (C), -0.1e (F), whereas the charges of the 1-butyl-1-methylpyrrolidinium cation were uniformly scaled. All proposed charges for all six ionic liquid are summarized in the FF files provided in the Supporting Information. Unlike in the pyridinium-based RTILs, the cation-anion closest-approach distance in the pyrrolidinium-based RTILs is described well by the existing CL&P FF parameters. Anions are coordinated by the methyl group, as opposed to coordination by the ring occurring in the pyridinium-based RTILs. This is, in our opinion, a principal difference between pyridinium-based and pyrrolidinium-based RTILs, which was not underlined in the literature thus far.

The refined FF treats all atoms explicitly, including hydrogen atoms. This is done for compatibility with the CL&P FF. Construction of the united atom force field is an interesting option, which would allow avoiding a number of dihedral functions in the Hamiltonian, and therefore, enhance sampling.

**Force Field Validation**

Figure 4 depicts temperature dependence of heat of vaporization, $H_{vap}$. All curves are nearly linear over the 298-353 K temperature range. In the case of molecular liquids, heat of vaporization is normally measured at the normal boiling point and then extrapolated to 298 K. The difference obtained after extrapolation is often inferior to the uncertainty of the measured value, which is high. The decrease of $H_{vap}$ with temperature increase is modest over the condensed matter temperature range. However, $H_{vap}$ decays drastically as temperature approaches critical point, which is not the case in the present study. The experimental $H_{vap}$ of [PY14][DCA] amounts to 161 kJ mol$^{-1}$ (Table S1), which is in a satisfactory agreement with the simulated $H_{vap}$ = 139 kJ mol$^{-1}$. The simulated $H_{vap}$ of [PY14][TFSI] equals to 150, whereas the experimental value is 152 kJ mol$^{-1}$ (Table S1), thus the model performs perfectly.

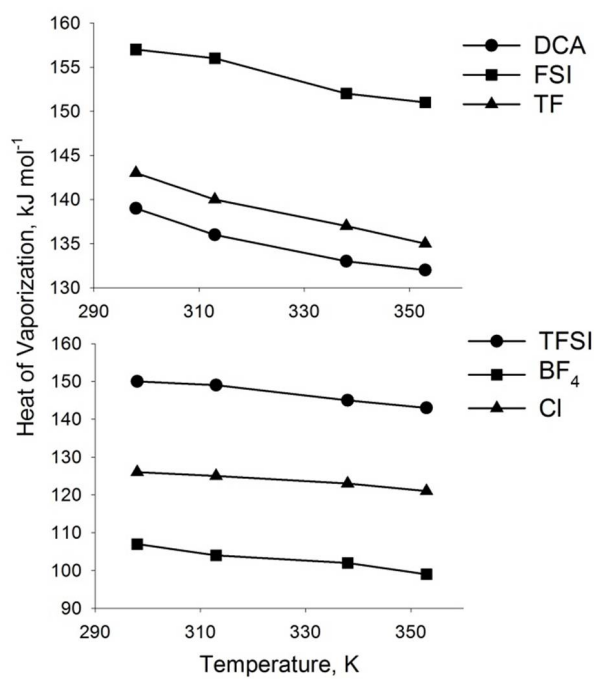

Figure 4. Heat of vaporization for the six simulated pyrrolidinium-based ionic liquids at 298, 313, 338, and 353 K.

Mass densities of the six pyrrolidinium-based RTILs (Figure 5) are in concordance with the experimental results (Table S1). Compare, 958 vs. 950 kg m$^{-3}$ for [PY14][DCA], 1326 vs. 1302 kg m$^{-3}$ for [PY14][FSI], 1398 vs. 1399 kg m$^{-3}$ for [PY14][TFSI]. The case of [PY14][TF] is the least successful out of these examples, 1204 vs. 1256 kg m$^{-3}$. The discrepancy of 4% is still within acceptable boundaries for classical molecular dynamics simulations. The reason of this discrepancy may be associated with less accurate atomic radii for specific cation-anion binding. [PY14][DCA] with a low density of 958 kg m$^{-3}$ at 298 K constitutes an interesting case. Less dense substances, in general, feature higher ionic conductivity and smaller shear viscosity. In turn, [PY14][TFSI] represents a notably dense ionic liquid thanks to relatively heavy sulfur and fluorine atoms in the anion.

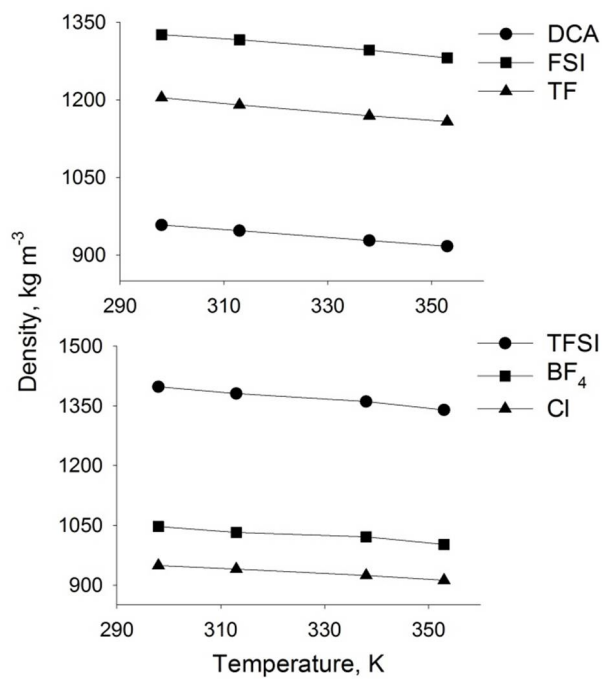

Figure 5. Mass density for the six simulated pyrrolidinium-based ionic liquids at 298, 313, 338, and 353 K. The size of circles was selected to be larger than the corresponding block-averaged standard errors (from simulations).

Indeed, both cation and anion of [PY14][DCA] exhibit higher self-diffusivities, 19 and $25 \times 10^{-11}$ m$^2$ s$^{-1}$ (Table 6) than any other ion from our list. These diffusions constants are in excellent agreement with the available experiments (Table S1). The slowest diffusion is observed in [PY14][FSI] and [PY14][TFSI], which is in line with their molecular masses and densities. In no RTIL, self-diffusion of cation differs from that of counter-ion significantly. Therefore, the case of very mobile cation and relatively slow anion or contrariwise is not observed. Small anions (chloride, tetrafluoroborate) coupled with [PY14]$^+$ become immobile due to a strong electrostatic attraction (high potential energies) in these systems.

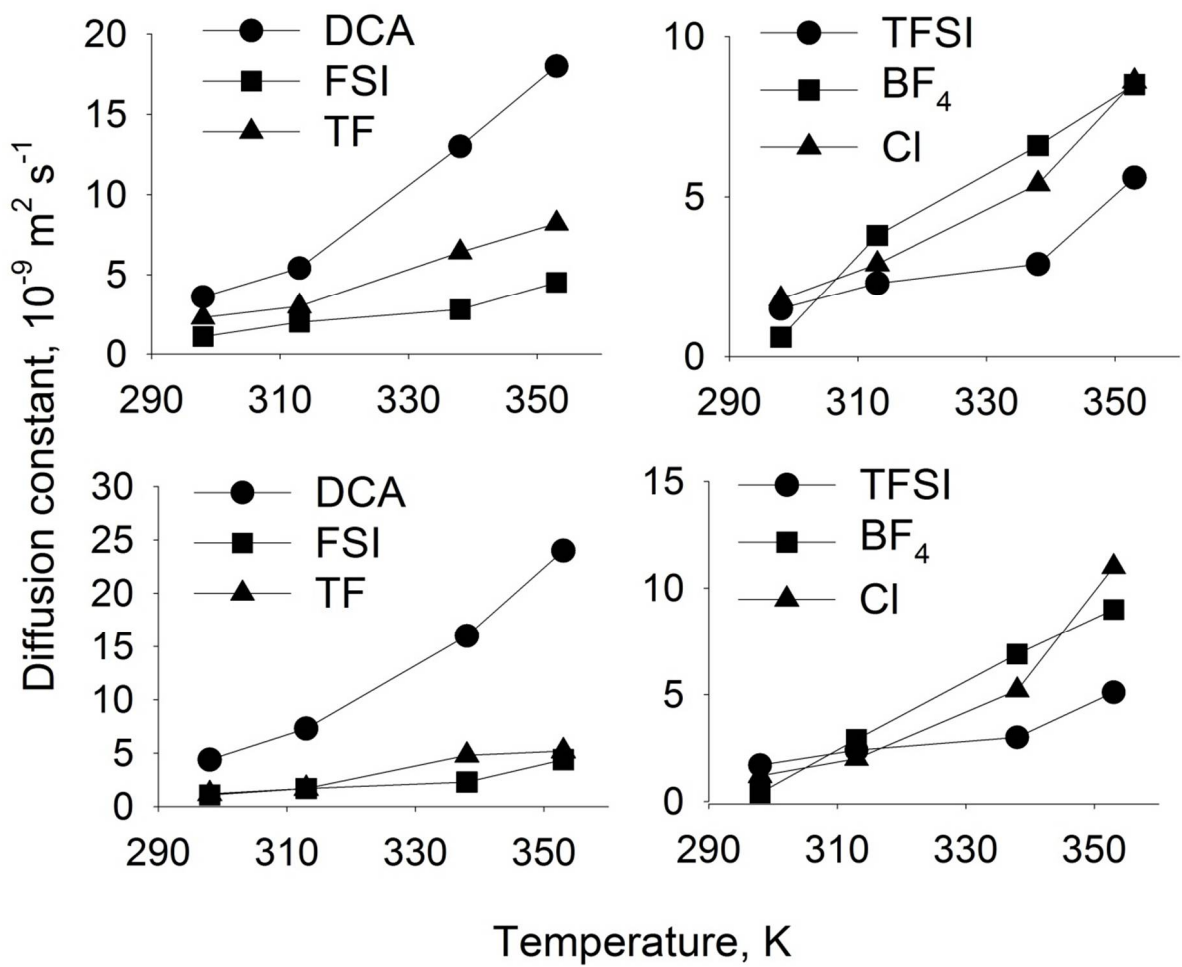

Figure 6. Self-diffusion constants for the six simulated pyrrolidinium ionic liquids at 298, 313, 338, and 353 K. The self-diffusion constants of the [PY14] cation are depicted in the upper panel, whereas the self-diffusion constants of the respective anions are depicted in the lower panel. The size of symbols is chosen to exceed standard errors of the computed self-diffusion constants at each temperature.

The simulated shear viscosities at 298, 313, 338, and 353 K are depicted in Figure 7. The comparison between experimental and simulated viscosities is provided in Figure 8. Exponential growth of viscosity with linear temperature increase was successfully achieved. The highest viscosities were observed for [PY14][TFSI] (118 cP) and [PY14][BF$_4$] (375 cP) at room temperature. In the first case, it predominantly occurs due to a large anionic species, whereas strong electrostatic contribution (polar covalent boron-fluorine bonds) is responsible for the second case. Interestingly, the chloride liquid is not most viscous in the current dataset. According to the developed force field and DFT calculations, the interaction between [Cl]$^-$ and

[PY14]$^+$ is not extremely strong. Indeed, Figure 1 suggests that [Cl]$^-$ is coordinated by the hydrocarbon (methyl) groups of the two neighboring pyrrolidinium-based cations, as opposed to coordination by the pyrrolidine ring. Unfortunately, experimental data are not yet available for the chloride containing RTIL.

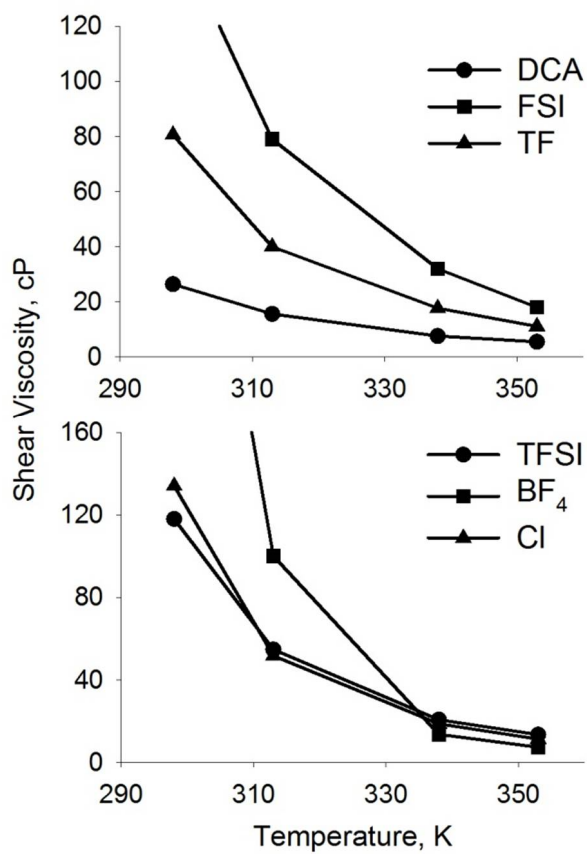

Figure 7. Shear viscosity for the six simulated pyrrolidinium-based ionic liquids at 298, 313, 338, and 353 K.

Correspondence between simulated and experimental data is best at 353 K (Figure 8). We have ensured that poorer performance of the developed FF at lower temperatures (298-338 K) is not due to insufficient sampling. This test has been done by computing viscosities from the subsequent parts of the trajectory and comparing them to one another. No drift was observed. The computed viscosities of [PY14][FSI] at all temperatures appear significantly and systematically higher than the experimental values (Table S1). The physical / computational reasons of such a behavior are not clear. It is especially surprising because other models (five

separate RTILs) of pyrrolidinium-based ILs exhibit an acceptable performance. Our FF derivation philosophy does not allow to. Therefore, we retain our FF parameters as they are. We must not exclude a possibility that the experimental viscosities are underestimated. Such a situation can occur, for instance, due to poor drying of this particular IL prior to viscosity measurements.

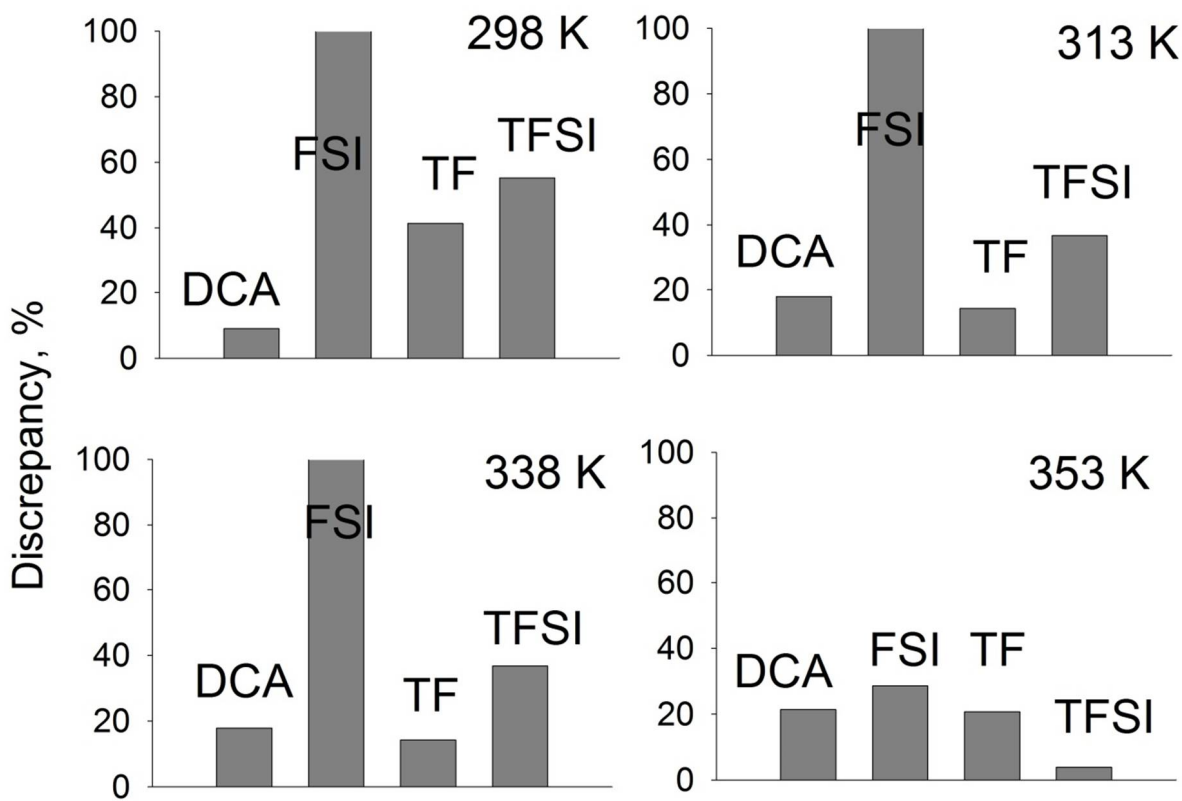

Figure 8. Divergence of the simulated shear viscosities and the experimental shear viscosities. The divergence (in pct.) is obtained as $\eta = \frac{\eta\_sim - \eta\_exp}{\eta\_exp} \times 100\%$.

The reported shear viscosities were computed using separate non-equilibrium MD simulations based on how fast extra energy is dissipated in the system. The deviation from equilibrium (degree of non-equilibrium) must be strictly controlled. Too small accelerations would result in unsatisfactorily slow sampling, whereas too large accelerations would lead to viscosity underestimation. To validate the acceleration value, which we chose, a range of alternative accelerations has been tested (Figure 9). Indeed, a significantly larger acceleration, such as 0.1 nm ps$^{-2}$, provides twice as smaller viscosity. The difference between 0.02 and 0.01 nm ps$^{-2}$ appears negligible; therefore, the chosen value provides a good balance between good accuracy and fast sampling. Note, that optimal acceleration depends simultaneously upon a liquid nature and an anticipated length of MD trajectory.

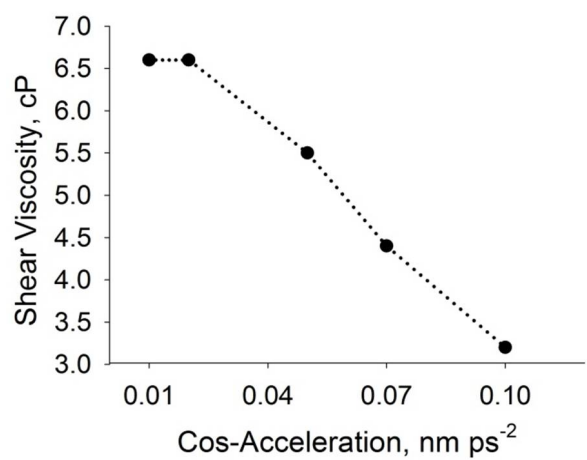

Figure 9. Shear viscosity computed using cosine-shape acceleration for various acceleration values: 0.01 … 0.10 nm ps$^{-2}$. Note, that the obtained viscosity must be extrapolated to zero acceleration. However, in practice the difference between small accelerations appears smaller than standard error of the computed viscosity. The reported viscosities of pyrrolidinium-based RTILs were calculated using an applied acceleration of 0.02 nm ps$^{-2}$.

**Conclusions**

Starting from the CL&P force field for the six pyrrolidinium-based ionic liquids, we applied a systematic refinement for more realistic reproduction of transport properties (ionic self-diffusion coefficients, shear viscosity). The improvement was achieved thanks to description of electrostatic potential of ion aggregates as opposed to isolated ions in vacuum (Lopes-Pádua force field). All properties of interest are in reasonable concordance with the available experimental values. Remarkably, the coincidence with the experimental properties is better in the case of anions, which do not contain sulfur. The likely reason is that sulfur, belonging to the third period of Periodic Table, exhibits significantly different polarizability, as compared to all other elements constituting these RTILs.

The developed force field fosters computational investigation of room-temperature ionic liquids and promotes understanding of physical and chemical atomistic-precision interactions in the condensed state of these complicated systems.

**Acknowledgments**

V.V.C. acknowledges a research grant from CAPES (Coordenação de Aperfeiçoamento de Pessoal de Nível Superior, Brasil) under "Science Without Borders" program.

**Supporting Information**

All force field parameters derived and tested in the manuscript can be found in Supporting Information. Table S1 summarizes all currently available experimental data on heat of vaporization, mass density, cationic and anionic self-diffusion constants, and shear viscosity at 298, 313, 338, and 353 K. This information is available free of charge via the Internet at http://pubs.acs.org.

**Contact Information**

E-mail for correspondence: vvchaban@gmail.com (V.V.C.); voroshylova@gmail.com (I.V.V.)

TOC Image

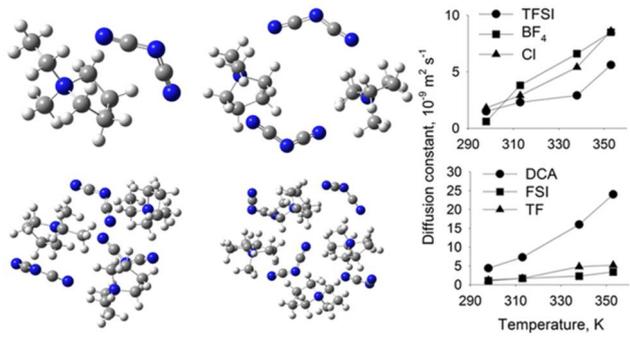